\documentclass[twoside]{ilcws07}
\usepackage[latin1]{inputenc}
\usepackage[dvips]{graphicx,epsfig,color}
\usepackage{wrapfig,rotating}
\usepackage{amssymb,amsmath,array}

\def\vecalpha{{\pmb\alpha}}

\begin{document}
\title{
%%%%   Paper title goes here  %%%%%%%%%%%%%%
Profile of Two-Higgs-Doublet-Model Parameter Space} %% 
%***********************************************************************
% AUTHORS INFORMATION AREA
%***********************************************************************
\author{Abdul Wahab El Kaffas$^1$, Odd Magne Ogreid$^2$ and Per Osland$^1$
% Optional short acknowledgment: remove next line if non-needed
%\thanks{This is an optional funding source acknowledgment.}
% DO NOT MODIFY THE FOLLOWING '\vspace' ARGUMENT
\vspace{.3cm}\\
% Addresses and institutions (remove "1- " in case of a single institution)
1- Department of Physics and Technology, University of Bergen \\
Postboks 7803, N-5020 Bergen, Norway
%% Remove the next three lines in case of a single institution
\vspace{.1cm}\\
2- Bergen University College,
Bergen, Norway\\
}
%%***********************************************************************
% END OF AUTHORS INFORMATION AREA
%***********************************************************************

\maketitle

\begin{abstract}
We review recent work on constraining the parameter space of the
Two-Higgs-Doublet Model by theoretical and experimental results. Some
characteristics of the model, in particular the distribution of masses
in the surviving parameter space, are discussed.
\end{abstract}

\section{Introduction}

We report on recent work on constraining the multi-dimensional parameter space
of the Two-Higgs-Doublet Model by theoretical and experimental results
\cite{url,Wahab El Kaffas:2007xd}.

As compared with the Standard Model (SM), the Two-Higgs-Doublet Model (2HDM)
allows for an additional mechanism for CP violation
\cite{Lee:1973iz}.  This is one of
the main reasons for continued strong interest in the model
\cite{Accomando:2006ga}.

Several experimental constraints restrict its parameter space.  The $B\to
X_s\gamma$ rate excludes low values of the charged-Higgs mass, $M_{H^\pm}$
\cite{Misiak:2006zs},
whereas $B-\bar B$ oscillations and the branching ratio $R_b$ for $Z\to b\bar
b$ exclude low values of $\tan\beta$.  The precise measurements at LEP of the
$\rho$ parameter constrain the mass splitting in the Higgs sector, and force
the masses to be not far from the $Z$ mass scale \cite{Bertolini:1985ia}.

From the theoretical point of view, there are also consistency conditions.
The potential has to be positive for large values of the fields
\cite{Deshpande:1977rw,ElKaffas:2006nt}.  Furthermore, we require the
tree-level Higgs--Higgs scattering amplitudes to be unitary
\cite{Kanemura:1993hm}.  Together, these constraints dramatically reduce the
allowed parameter space of the model.  In particular, the unitarity constrain
excludes large values of $\tan\beta$, {\it unless} $\mu$ is reasonably
large. This limit is basically the decoupling limit \cite{Gunion:2002zf}.

Our recent study \cite{Wahab El Kaffas:2007xd}, restricted to the so-called
``Type II'' version, where up-type and down-type quarks couple to different
Higgs doublets, uses rather complete and up-to-date experimental results, as
well as accurate theoretical predictions for the above quantities.  We
consider a model with the $Z_2$ symmetry respected by the quartic couplings,
i.e., no $\lambda_6$ and $\lambda_7$ couplings.  Otherwise, we allow for full
generality. In particular, we allow for $CP$ violation, taking $\lambda_5$
complex.  (For a definition of the potential, see \cite{Wahab El
Kaffas:2007xd}.)  The neutral Higgs boson sector will thus contain three
bosons, described by a $3\times3$ mixing matrix $R$.  These three neutral
Higgs bosons will in general all have $CP$-violating Yukawa couplings.
A related study, focused more on large values of $\tan\beta$, was also
presented at this Workshop \cite{Krawczyk}.

\section{Results}

We parametrize the model in terms of the masses of the two lightest 
neutral Higgs bosons, together with the charged Higgs boson mass, $\tan\beta$,
the soft parameter $\mu^2$, and the rotation matrix $R$ of the neutral sector.
The third (heaviest) neutral mass is then calculable, as well as the quartic
couplings, $\lambda_i$ (see \cite{Khater:2003wq,Kaffas:2007rq}).

\begin{wrapfigure}{r}{0.68\columnwidth}
\centerline{\includegraphics[width=0.75\columnwidth]{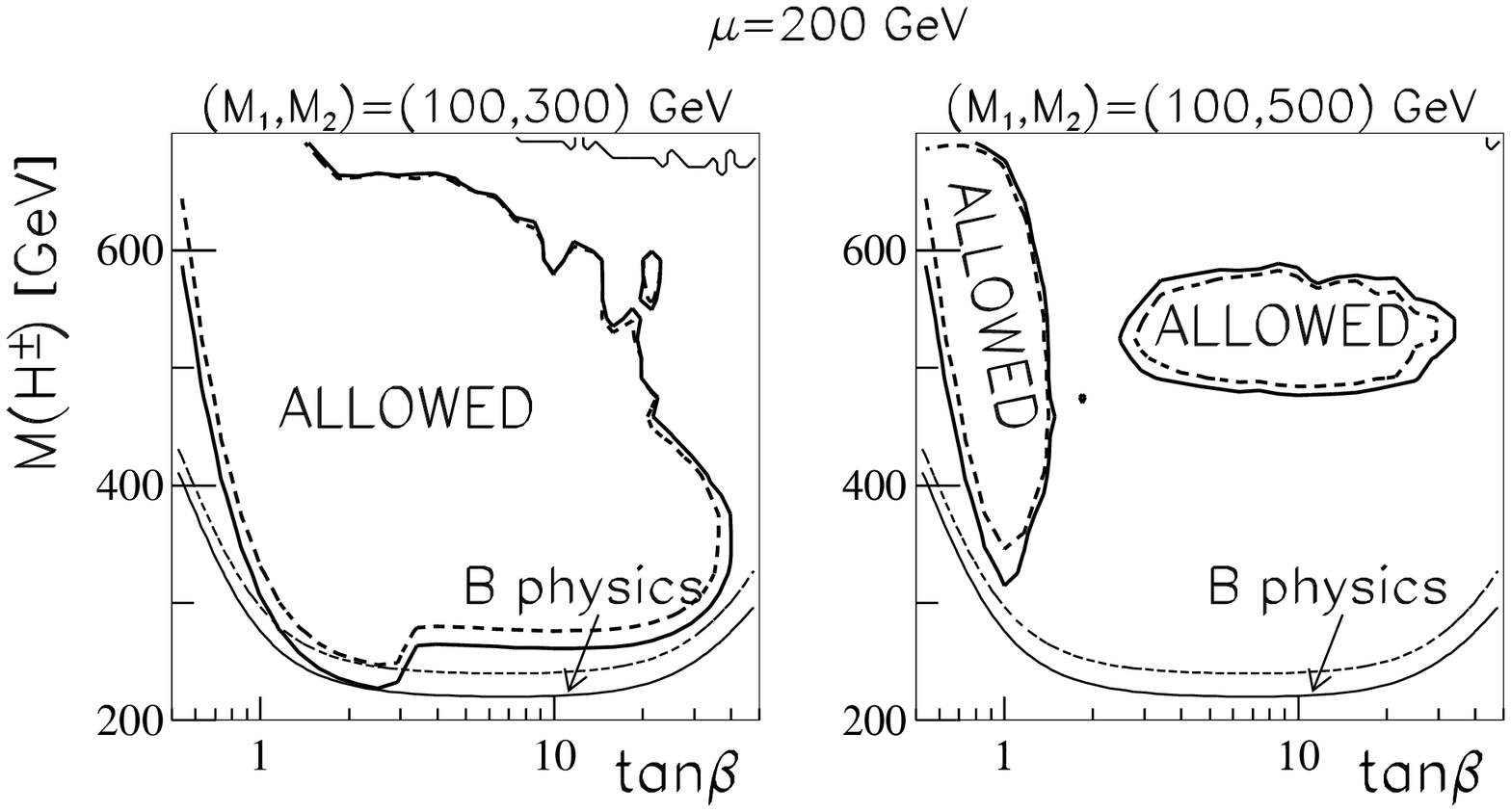}}
\centerline{\includegraphics[width=0.75\columnwidth]{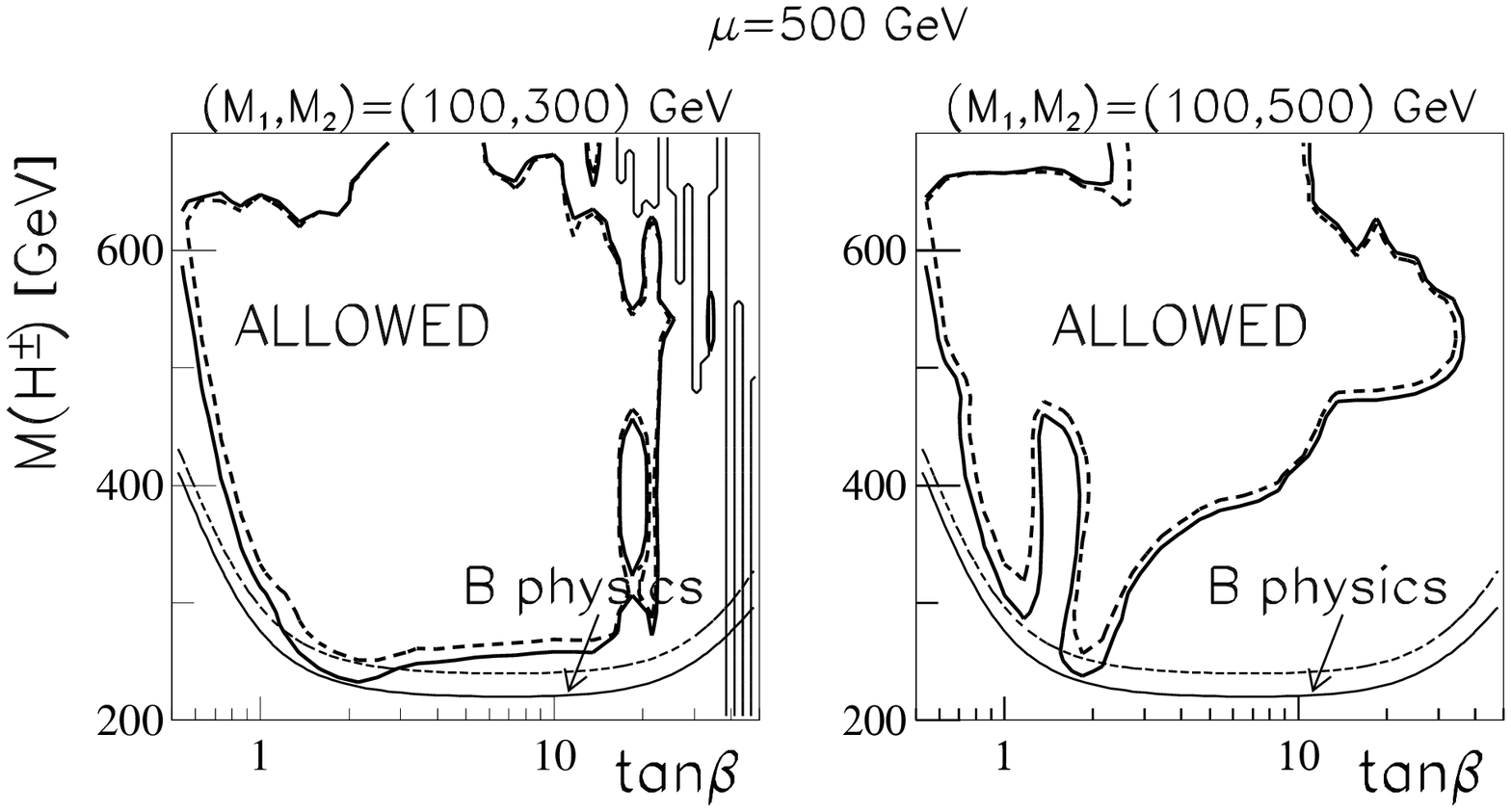}}
\caption{Allowed regions in the $\tan\beta$--$M_{H^\pm}$ plane,
taking into account theoretical and experimental constraints.}
\label{Fig:allowed}
\end{wrapfigure}

We establish allowed regions in the $\tan\beta$--$M_{H^\pm}$ plane
by the following procedure: For each point in this plane, we scan
over the parameters $\vecalpha=\{\alpha_1,\alpha_2,\alpha_3\}$,
defining the mixing matrix $R$ in the neutral-Higgs sector,
imposing the absolute theory constraints of positivity and unitarity.
At each point, we evaluate a $\chi^2$ penalty corresponding to the
experimental constraints, adopting the ``best'' point (lowest $\chi^2$)
in $\vecalpha$.

For two values of $\mu$ (200 and 500~GeV), we show in Fig.~\ref{Fig:allowed}
the allowed regions in the $\tan\beta$--$M_{H^\pm}$ plane, taking into account
the theoretical constraints mentioned above, the LEP2 non-discovery, the very
precise $\Delta\rho$ measurements at LEP, as well as the $B$-physics
constraints ($B\to X_s\gamma$, mainly), and $R_b$.  The masses of the two
lightest neutral Higgs bosons are here kept fixed, at $M_1=100~\text{GeV}$ and
$M_2=300~\text{GeV}$ or 500~GeV.

The over-all surviving regions of parameter space depend significantly on
the ``soft'' parameter $\mu^2$.
At low or negative values, the unitarity constraint will cut off
the allowed region already at moderate values of $\tan\beta$.
We have therefore shown results for a couple of positive values
of $\mu^2$, the higher one approaching the so-called decoupling limit.

\section{Distribution of Higgs masses}

\begin{figure}
\centerline{\includegraphics[width=\columnwidth]{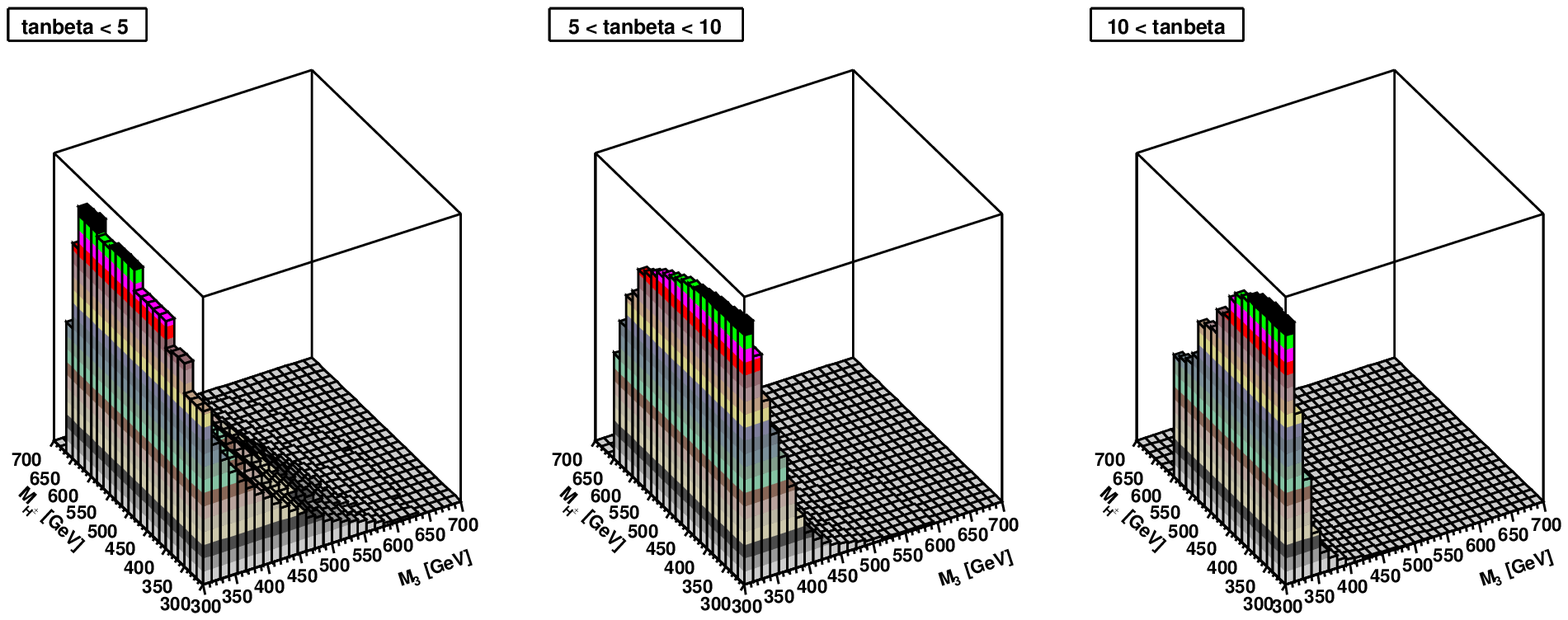}}
\centerline{\includegraphics[width=\columnwidth]{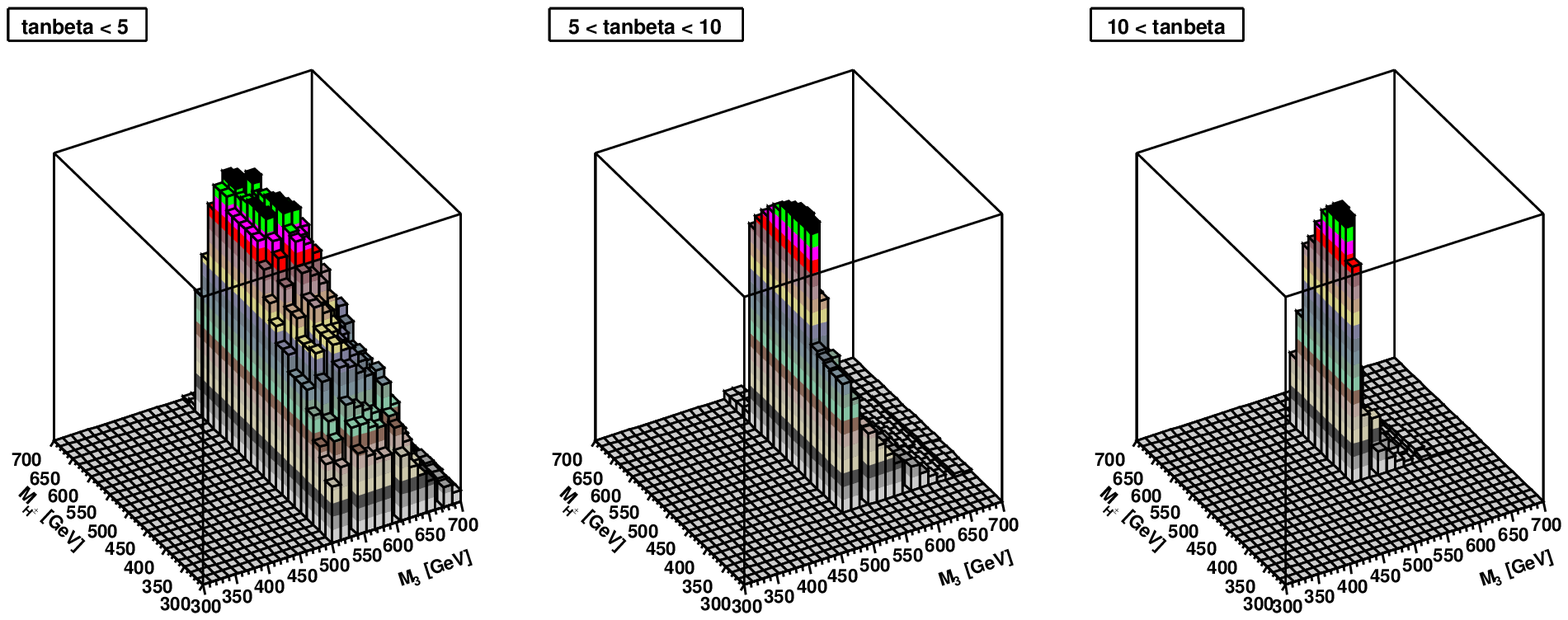}}
\caption{Distribution of $M_3$-values for fixed $M_1=100~\text{GeV}$.
Top:
$M_2=300~\text{GeV}$ and $\mu=200~\text{GeV}$;
bottom:
$M_2=500~\text{GeV}$ and $\mu=500~\text{GeV}$. Three slices of
$\tan\beta$-values are shown.}
\label{Fig:M3-values-low-mu}
\end{figure}

It turns out that, if $\mu$ is comparable with $M_2$, or smaller, the
distribution of $M_3$-values will be very narrow, especially at large values
of $\tan\beta$. This is illustrated in Fig.~\ref{Fig:M3-values-low-mu}, for
$M_1=100~\text{GeV}$, and two sets of $(M_2,\mu)$ values:
$(300,200)~\text{GeV}$ and $(500,500)~\text{GeV}$.  Also, we note that for
$M_2=500~\text{GeV}$ and $\mu=500~\text{GeV}$ (lower panels), low values of
$M_{H^\pm}$ are excluded.  This is basically because of the $\Delta\rho$
constraint.

\begin{figure}
\centerline{\includegraphics[width=\columnwidth]{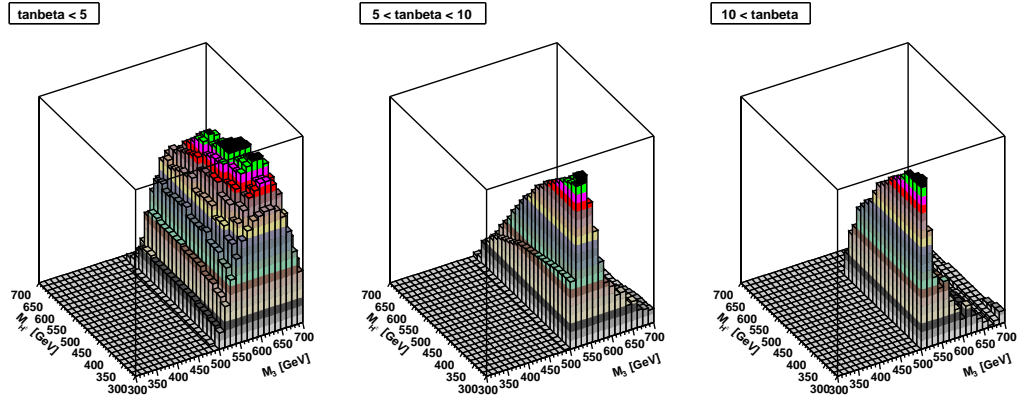}}
\caption{Distribution of $M_3$-values for $M_1=100~\text{GeV}$,
$M_2=300~\text{GeV}$ and $\mu=500~\text{GeV}$. Three slices of
$\tan\beta$-values are shown, increasing to the
right.}\label{Fig:M3-values-high-mu}
\end{figure}

On the other hand, if $\mu$ is larger than $M_2$, the distribution can be 
considerably wider, as is seen in Fig.~\ref{Fig:M3-values-high-mu}.

\section{Summary}

We have shown that the 
constraints of positivity and tree-level
unitarity of Higgs--Higgs scattering,
$B$-physics results, together with the precise
LEP measurements, in particular of the $\rho$-parameter at LEP,  
exclude large regions of the 2HDM~(II)
parameter space.  High values of $\tan\beta$ are excluded unless $\mu$ is
large, allowing $M_2$ and $M_3$ both to be
heavy. Furthermore, $M_2$ and $M_3$ should be reasonably close to each
other.  Improved precision of the $\bar B\to X_s\gamma$ measurement could
significantly reduce the remaining part of the parameter space, but it appears
unlikely that the model could be excluded other than by a negative search at
the LHC. 

\section*{Acknowledgments}

This research has been supported in part by
the Mission Department of Egypt and the Research Council of Norway.

% ****************************************************************************
% BIBLIOGRAPHY AREA
% ****************************************************************************

\begin{footnotesize}
% IF YOU DO NOT USE BIBTEX, USE THE FOLLOWING SAMPLE SCHEME FOR THE REFERENCES
% ----------------------------------------------------------------------------

\end{footnotesize}

% ****************************************************************************
% END OF BIBLIOGRAPHY AREA
% ****************************************************************************

\end{document}